\documentclass[12pt,showpacs]{revtex4} 
\usepackage{amssymb,epsf}
\usepackage{latexsym}
\usepackage{color}
\begin{document}

\title{\textbf{Power-law solutions in $f(T)$ gravity}}
\author{M. R. Setare}
\email{rezakord@ipm.ir}
\affiliation{Department of Science\\ Payame Noor University, Bijar, Iran}
\author{F. Darabi}
\email{f.darabi@azaruniv.edu} 
\affiliation{Department of Physics\\ Azarbaijan University of Shahid Madani, Tabriz 53714-161, Iran \\
Research Institute for Astronomy and Astrophysics of Maragha (RIAAM),
Maragha 55134-441, Iran\\
Corresponding Author}

\date{\today}
\begin{abstract}
\textbf{Abstract:} 
We have considered an action of the form $T+f(T)+L_m$ 
describing Einstein's gravity plus a function of the torsion
scalar. By considering an exact power-law solution we have obtained the Friedmann equation as a differential equation for the function $f(T)$ in spatially flat universe and obtained the real valued solutions of this equation for some power-law solutions. We have also studied the power-law solutions when the universe enters a Phantom phase and shown that such solutions
may exist for some f(T) solutions.


\end{abstract}
\maketitle
\newpage
\section{Introduction}
Recent cosmological observations indicate that our universe is in
accelerated expansion. These observations are those which is
obtained by SNe Ia {\cite{c1}}, WMAP {\cite{c2}}, SDSS {\cite{c3}}
and X-ray {\cite{c4}}.  These observations also suggest that our
universe is spatially flat, and consists of about $70 \%$ dark
energy (DE) with negative pressure, $30\%$ dust matter (cold dark
matter plus baryons), and negligible radiation. In order to
explain why the cosmic acceleration happens, many theories have
been proposed. The simplest candidate of the dark energy is a tiny
positive time-independent cosmological constant $\Lambda$, for
which $\omega=-1$. However, it is difficult to understand why the
cosmological constant is about 120 orders of magnitude smaller
than its natural expectation (the Planck energy density). This is
the so-called cosmological constant problem. Another puzzle of
the dark energy is the cosmological coincidence problem: why are
we living in an epoch in which the dark energy density and the
dust matter energy are comparable?. An alternative proposal for
dark energy is the dynamical dark energy scenario.
 The dynamical nature of dark
energy, at least in an effective level, can originate from various
fields, such is a canonical scalar field (quintessence)
\cite{quint}, a phantom field, that is a scalar field with a
negative sign of the kinetic term \cite{phant}, or the combination
of quintessence and phantom in a unified model named quintom
\cite{quintom}. Recently another paradigm has been constructed in
the light of the holographic principle of quantum gravity theory,
and thus it presents some interesting features of an underlying
theory of dark energy \cite{holoprin}. This paradigm may
simultaneously provide a solution to the coincidence problem
\cite{Li:2004rb}.
\\
It is known that Einstein's theory of gravity may not describe
gravity at very high energies. The simplest alternative to
general relativity is Brans-Dicke scalar-tensor theory \cite{10}.
Modified gravity also provides the natural gravitational
alternative for dark energy \cite{11}. Moreover, thanks to the
different roles of gravitational terms relevant at small and at
large curvature, the modified gravity presents natural
unification of the early-time inflation and late-time
acceleration . It may naturally describe the transition from
non-phantom phase to phantom one without necessity to introduce
the exotic matter. But among the most popular modified gravities
which may successfully describe the cosmic speed-up is $f(R)$
gravity. Very simple versions of such theory like $\frac{1}{R}$
\cite{12} and $\frac{1}{R}+R^2$ \cite{13} may lead to the
effective quintessence/phantom late-time universe (to see solar
system constraints on modified dark energy models refer to
\cite{14}, also general review of reconstruction is given in
\cite{noji} ). Another theory proposed as gravitational dark energy
is scalar-Gauss-Bonnet gravity $f(G)$ \cite{15} which is closely
related with the low-energy string effective action. In this
proposal, the current acceleration of the universe may be caused
by mixture of scalar phantom and (or) potential/stringy effects.
On the other hand, a theory of $f(T)$ gravity has recently been
received attention. Models based on modified teleparallel gravity
were presented, in one hand, as an alternative to inflationary
models \cite{16, 17}, and on the other hand, as an alternative to
dark energy models \cite{18}. In this paper, we show a
cosmological power law solution for the acceleration of the
universe based on the above mentioned modification of the
teleparallel equivalent of General Relativity.
We consider $T+f(T)$ gravity model and reconstruct this
theory from the cosmological power law solution for the scale
factor. We know that the power law solutions are very important in
the standard cosmology, because this type of solutions provides a framework for establishing the behaviour of more general cosmological solutions in
different histories of our universe, such as radiation dominant, matter dominant
or dark energy dominant eras.

\section{Field equations for  $[T+f(T)+L_m]$ gravity }

The action for the theory of modified gravity based on a
modification of the teleparallel equivalent of General
Relativity, namely $f(T)$ theory of gravity, coupled with matter
$L_m$ is given by \cite{18}, \cite{19} and \cite{20}
\begin{equation}\label{1}
S=\frac{1}{16\pi G}\int d^4x e \left[T+f(T)+L_m\right]_,
\end{equation}
where $e=det(e^i_{\mu})=\sqrt{-g}$. The teleparallel Lagrangian
$T$ is defined as follows
\begin{equation}\label{2}
T=S^{\:\:\:\mu \nu}_{\rho} T_{\:\:\:\mu \nu}^{\rho},
\end{equation}
where
$$
T_{\:\:\:\mu \nu}^{\rho}=e_i^{\rho}(\partial_{\mu}
e^i_{\nu}-\partial_{\nu} e^i_{\mu}),
$$
$$
S^{\:\:\:\mu \nu}_{\rho}=\frac{1}{2}(K^{\mu
\nu}_{\:\:\:\:\:\rho}+\delta^{\mu}_{\rho} T^{\theta
\nu}_{\:\:\:\theta}-\delta^{\nu}_{\rho} T^{\theta
\mu}_{\:\:\:\theta}),
$$
and $K^{\mu \nu}_{\:\:\:\:\:\rho}$ is the contorsion tensor
$$
K^{\mu \nu}_{\:\:\:\:\:\rho}=-\frac{1}{2}(T^{\mu
\nu}_{\:\:\:\:\:\rho}-T^{\nu \mu}_{\:\:\:\:\:\rho}-T^{\:\:\:\mu
\nu}_{\rho}).
$$
The field equations are obtained by varying the action with
respect to vierbein $e^i_{\mu}$ as follows
\begin {equation}\label{3}
e^{-1}\partial_{\mu}(e S^{\:\:\:\mu
\nu}_{i})(1+f_T)-e_i^{\:\lambda}T_{\:\:\:\mu
\lambda}^{\rho}S^{\:\:\:\nu \mu}_{\rho}f_T +S^{\:\:\:\mu
\nu}_{i}\partial_{\mu}(T)f_{TT}-\frac{1}{4}e_{\:i}^{\nu}
(1+f(T))=4 \pi G e_i^{\:\rho}T_{\rho}^{\:\:\nu},
\end{equation}
where $f_T=f'(T)$ and $f_{TT}=f''(T)$. Now, we take the usual
spatially-flat metric of Friedmann-Robertson-Walker (FRW)
universe, in agreement with observations
\begin {equation}\label{4}
ds^{2}=dt^{2}-a(t)^{2}\sum^{3}_{i=1}(dx^{i})^{2},
\end{equation}
where $a(t)$ is the scale factor as a one-parameter function of
the cosmological time $t$. Moreover, we assume the background to
be a perfect fluid. Using the Friedmann-Robertson-Walker metric
and the perfect fluid matter in the teleparallel Lagrangian
(\ref{2}) and the field equations (\ref{3}), one obtains
\begin {equation}\label{4'}
T=-6H^2,
\end{equation}
\begin {equation}\label{5}
H^2=\frac{8 \pi G \rho}{3}-\frac{1}{6}f-2H^2f_T,
\end{equation}
\begin {equation}\label{5'}
\dot{H}=-\frac{4 \pi G (\rho+p)}{1+f_T-12H^2f_{TT}},
\end{equation}
where $\rho$ and $p$ denote the matter density and pressure
respectively, and the Hubble parameter $H$ is defined by
$H=\dot{a}/a$.

In the FRW universe, the energy conservation law can be expressed
as the standard continuity equation
\begin {equation}\label{6}
\dot{\rho}+3H(\rho+p)=\dot{\rho}+3H(1+w)\rho=0\ ,
\end{equation}
where $\rho$ is the matter energy density and $p=w\rho$ is the
equation of state relating pressure $p$ with energy density. 

\section{Exact matter dominant power-law solutions }

We now assume an exact power-law solution for the field equations
\begin{equation}\label{9}
a(t)=a_0 t^m,
\end{equation}
where $m$ is a positive real number. From
the assumption (\ref{9}) and the continuity equation (\ref{6}), we obtain
\begin {equation}\label{7}
\rho(t)=\rho_{0}t^{-3m(1+w)}.
\end{equation} Moreover, using the assumption (\ref{9}),
Eq.(\ref{4'}) leads us to the following result
\begin {equation}\label{10}
T=-6\frac{m^2}{t^2}<0.
\end{equation}
By using Eqs.(\ref{4'}), (\ref{7}) and (\ref{10}) in Eq.(\ref{5}), we obtain
the Friedmann equation
\begin {equation}\label{11}
\frac{T}{3}f_T+\frac{1}{6}(T-f)+\frac{8 \pi G
}{3}\rho_{0}m^{-3m(1+w)}(-\frac{1}{6}T)^{\frac{3}{2}m(1+w)}=0.
\end{equation}
This is a differential equation for the function $f(T)$. The
general solution of this equation is obtained as
\begin{equation}\label{12}
f \left( T \right)=C_1\sqrt{T}-
\frac{2^{4-\frac{3}{2}m(1+w)}(3)^{-\frac{3}{2}m(1+w)}m^{-3m(1+w)}(-T)^{\frac{3}{2}m(1+w)}}{3m(1+\omega)-1}\pi
G \rho_0-T ,
\end{equation}
where $C_{1}$ is an arbitrary constant of integration. Because of $T<0$, in order to avoid of imaginary function $f(T)$, with no loss of generality
we may assume the constant $C_{1}=0$. Hence, the function $f(T)$ becomes \begin{equation}\label{13} f \left( T\right)=\frac{2^{4-\frac{3}{2}m(1+w)}(3)^{-\frac{3}{2}m(1+w)}m^{-3m(1+w)}(-T)^{\frac{3}{2}m(1+w)}}{1-3m(1+\omega)}\pi
G \rho_0-T.
\end{equation}
We note that having a finite real valued solution of $f(T)$
requires also one of the followings:\\
1) $m>0$ for any value of $\omega$, \\
2) $m<0$ for $-3m(1+\omega)\neq$ {half integer},\\
3) $m=0$ ,\\
all subject to the condition $3m(1+\omega)\neq1$. While the first
case leads to an expanding universe, the second case describes a
contracting universe, and the third case describes an static
universe. Therefore, power law solutions exist for the function
(\ref{13}) subject to the first and second conditions above.

By inserting (\ref{13}) in the action (\ref{1}), we find that the
standard Einstein gravity will automatically be recovered when
$f(T)=0$. This happens provided that ${\frac{3}{2}m(1+w)}=1$ and
\begin {equation}\label{14}
\rho_0=\frac{3m^{3m(1+w)}}{8\pi G }=\frac{3m^{2}}{8\pi G },
\end{equation}
which guarantees the positivity requirement of $\rho_0$.

\section{Exact Phantom phase power-law solutions }

One may also study the power-law solutions where the universe
enters a phantom phase leading to a Big Rip singularity. For this
case, the general class of Hubble parameters and cosmological
solutions are defined as
\begin {equation}\label{20}
H(t)=\frac{m}{t_s-t},
\end{equation}
\begin {equation}\label{20'}
a(t)=a_0 (t_s-t)^{-m},
\end{equation}
where $t_s$ is the so called ``Rip time'' at future singularity.
It is easy to show that all the results in sec IV follow from sec II by replacing m by -m.

Demanding a Big Rip during the phantom phase, as the cosmic time $t$
approaches $t_s$, requires $m\geq1$ in (\ref{20'}). Therefore,
power law solutions for the Phantom phase exist for the corresponding function
$f(T)$.

\section{Conclusion}

In the present paper we have considered a $T+f(T)+L_m$ action
which describes Einstein's gravity plus a function of the torsion
scalar. Then, by considering an exact power-law solution for the
field equations we have obtained the Friedmann equation in
spatially flat universe. The Friedmann equation appears as a
differential equation for the function $f(T)$. We obtained the
solution of this equation and showed that our model with this
solution for $f(T)$ has power-law solution of the type $a(t)=a_0
t^m$. We have also studied the power-law solutions when the
universe enters a Phantom phase. By considering such power-law
solution for the field equations, the corresponding
Friedmann equation and the solution $f(T)$ is simply obtained by comparing
with the results obtained in non-Phantom phase and replacing $m$ by $-m$. It is shown that the power-law solution of the type $a(t)=a_0(t_s-t)^{-m}$ also exists in
the phantom phase for this $f(T)$ solution.

\section*{Acknowledgment}
This work has been supported financially by Research Institute for Astronomy and Astrophysics of Maragha (RIAAM) under research project No. 1/2360.

\end{document}